# Superconductivity in $HfTe_5$ Induced via Pressures

Y. Liu[1*], Y. J. Long[1*], L. X. Zhao[1], S. M. Nie[1], S. J. Zhang[1], Y. X. Weng[1], M.L.Jin[1], W.M.Li[1], Q. Q. Liu[1], Y.W.Long[1], R.C.Yu[1], X. L. Fen[3], Q. Li[3],H.M.Weng[1,2], X.Dai[1,2], Z.Fang[1,2], G. F. Chen[1,2], C. Q. Jin[1,2]

1. Institute of Physics, Chinese Academy of Sciences, Beijing 100190, China
2. Collaborative Innovation Center of Quantum Matter, Beijing, China
3. Department of Materials Science, Jilin University, Changchun 130012, China

Recently, $ZrTe_5$ and $HfTe_5$ are theoretically studied to be the most promising layered topological insulators since they are both interlayer weakly bonded materials and also with a large bulk gap in the single layer. It paves a new way for the study of novel topological quantum phenomenon tuned via external parameters. Here, we report the discovery of superconductivity and properties evolution in $HfTe_5$ single crystal induced via pressures. Our experiments indicated that anomaly resistance peak moves to low temperature first before reverses to high temperature followed by disappearance which is opposite to the low pressure effect on $ZrTe_5$. $HfTe_5$ became superconductive above ~5.5 GPa up to at least 35 GPa in the measured range. The highest superconducting transition temperature (Tc) around 5 K was achieved at 20 GPa. High pressure Raman revealed that new modes appeared around pressure where superconductivity occurs. Crystal structure studies shown that the superconductivity is related to the phase transition from *Cmcm* structure to monoclinic *C2/m* structure. The second phase transition from *C2/m* to *P-1* structure occurs at 12 GPa. The combination of transport, structure measurement and theoretical calculations enable a completely phase diagram of $HfTe_5$ under high pressures.

*These authors contribute equally to this work.

To Whom Address: G.F.C.(gfchen@iphy.ac.cn) or C.Q.J. (Jin@iphy.ac.cn)

## Introduction

The topological quantum matters such as topological insulators(TIs), Dirac semimetals and Weyl semimetals are extensively studied recently as new frontier of condensed matter physics[1-11]. In 2D TIs, the electron moves without dissipation along edges due to time-reversal symmetry protection [12]. This ignited experimental follow up, such as the growth of stable 2D TIs [13,14]. On the path to find the new suitable 2D TIs, the single-layer $HfTe_5$ with large energy gap was proposed in the recent theoretical research [15]. Previously, the $HfTe_5$ has been studied as thermoelectric material that exhibits a resistivity anomaly at Tp~80K [16] where the sign of its thermoelectric power changes [17], showing a large magnetoresistance [18]. $HfTe_5$ is isostructural with $ZrTe_5$ [19] and has ellipsoidal anisotropic Fermi surface similar to that of $ZrTe_5$ [20]. Recent experimental study reported chiral magnetic effect [21] and quasi-2D Dirac fermions [22] in $ZrTe_5$, though it has no necessary crystal symmetry to protect the four-fold degenerate Dirac nodes [6]. However, the theoretical calculations predicate that both $ZrTe_5$ and $HfTe_5$ are located in the vicinity of a transition between strong and weak TI [15]. These interesting results make $HfTe_5$ a potential material for the study of the novel topological quantum phenomenon and topological phase transitions.

High pressure is a neat method being powerful to tune the electronic states and lattice structures of the quantum matters without introducing disorder or impurity inherent to chemical doping. In this work, we report the discovery of pressure induced superconductivity in $HfTe_5$ single crystals. Transport experiments indicate consecutive transitions induced by pressure from semimetal to metal before superconductivity appears at a critical pressure of ~5.5 GPa. Combined with theoretical calculations, the structure evolution as function of pressure is investigated by using Raman, which indicates that the superconductivity is correlated to the structural phase transition. Furthermore, a systematic phase diagram on crystal and electronic properties of $HfTe_5$ as a function of pressure is presented.

## Experiments

Single crystals of $HfTe_5$ were grown by chemical vapor transport. Stoichiometric amounts of Hf (powder, 3N, Zr nominal 3%) and Te (powder, 5N) were sealed in a quartz ampoule with iodine (7mg/mL). Quartz ampoule was placed in a two-zone furnace for almost one month with typical temperature gradient from 500℃ to 400℃ applied. $HfTe_5$ single crystals present long ribbon shape[23]. The crystal structure of $HfTe_5$ has been determined by powder X-ray diffraction experiments[20], which is orthorhombic with space group of *Cmcm* as shown in Figure S1. Trigonal prismatic chains of $HfTe_3$ run along the *a* axis, and these prismatic chains are linked via parallel zigzag chains of Te atoms along the c axis to form a 2D sheet of $HfTe_5$ in the *ac* plane. The sheets of $HfTe_5$ stack along the b axis, forming a layered structure [15].

The electronic transport properties of $HfTe_5$ single crystals at high pressure was measured using the standard four-probe method by diamond anvil cell (DAC) made of nonmagnetic BeCu alloy as described in [24-28]. Pressure was generated by a pair of diamonds with a 500mm diameter culet. A T301 stainless steel gasket, pre-indented from 250μm to 40μm thickness, was drilled a center hole with 250μm in diameter. The gasket was then covered by cubic BN fine powders to protect the electrode leads from the gasket. A center hole with a diameter of 100μm which serves as sample chamber was further drilled at the insulating layer. The $HfTe_5$ single crystal with a dimension of 80μm*80μm*10μm was loaded with soft NaCl fine powder surrounding as pressure transmitting medium. Slim gold wires of 18mm in diameter were used as electrodes. Pressure was calibrated by ruby fluorescence shift method for all the experiments. The DAC was placed inside a MagLab system to perform the electronic transport experiments. To ensure equilibrium, the MagLab system automatically controlled the temperature so that temperature slowly decreased. A thermometer located around the sample in the diamond anvil cell was used for monitoring the sample temperature.

The structure search simulations are performed through the CALYPSO method, which is specially designed for global structural minimization unbiased by any known structural information. The first principles calculations have been carried out by using the projector augmented wave (PAW) method implemented in *Vienna ab initio* *simulation package* (VASP). The lattice parameters determined by X-ray diffraction are adopted in our calculations. Generalized gradient approximation (GGA) of Perdew-Burke-Ernzerhof type is used. The k-point sampling grids are set to 14*14*8, 11*11*7 and 11*7*3 for the self-consistent calculations of $HfTe_5$ in 0 GPa, 10 GPa and 20 GPa, respectively. The cut-off energy for the plane wave expansion is chosen as 500 eV. Spin-orbit coupling (SOC) is taken into account self-consistently.

The high pressure Raman experiments are conducted on $HfTe_5$ single crystal at room temperature using Renishaw inVia Raman microscope system with laser wavelength 532 nm. The gaskets made of T301 stainless steel, are preindented from 250μm to 40μm and drilled a center hole of 120mm in diameter. In order to provide a good quasi-hydrostatic pressure environment, the $HfTe_5$ single crystal with a dimension of 60μm*60μm*10μm was loaded with daphne 7373 oil as pressure transmitting medium. Pressure was also calibrated by ruby fluorescence shift.

**Results and Discussions**

Figure 1 shows the evolution of *ac* plane resistance as a function of temperature of $HfTe_5$ single crystals at various pressures. At 1.3 GPa, the resistance displays a typical semiconductor-like behavior above 40K. As temperature continues to decrease, the resistance increases much slowly. When pressure increases up to 2.1 GPa, the resistance shows a hump near 49K, and then decreases with temperature, accompanied by an upturn below 11K. The behaviors of the abnormal resistance appearing at 40K and 49K are intimately tied to topological critical point, which are similar to those observed at ambient pressure[16,17,23]. The temperature of maximum resistance (Tp) increases to 84K at 4.0 GPa, accompanied by the broadening of the hump and the decrease of the peak resistance. Up to 5.5 GPa, in addition to the increases of Tp up to 136K, a small drop of resistance is observed at low temperature which indicates the occurrence of superconducting.

Comparing with the electronic transport properties of $ZrTe_5$ in previous report [29], both $HfTe_5$ and $ZrTe_5$ display a resistive abnormal hump. The Tp in $HfTe_5$ crystal decreases with the pressure up to 1.7GPa but those of $ZrTe_5$ is on the opposite. One explanation of this difference could be that the Fermi level is on the opposite sides of a peak in the density of states, thus pressure could reduce Tp in one case but increase Tp in the other case. This explains why the Tp of $HfTe_5$ under low pressure is lower than that at ambient pressure (80K). While keep increasing pressure, our experiment indicates that Tp changed along with pressure, which implies anomaly resistance peak moves to low temperature first before reverses to high temperature followed by disappearance. That is also opposite to the pressure effect on $ZrTe_5$[30].

Further increasing the pressure, the maximum of resistance is totally suppressed and the overall resistance shows a metallic transport behavior. A clear superconducting transition was observed at 5.5 GPa, as shown in Figure 1(b). The signature of superconductivity is the resistance drop at around 2.7 K. The transition temperatures was defined based on the differential of resistance over temperature (dR/dT)[24]. With the pressure increasing to 6.6 GPa, Tc grows rapidly with the drop of resistance getting more pronounced and the zero resistance starting to be fully realized. The superconductivity transition at pressures up to 35 GPa are shown in Figure 1(b). In the whole range of pressure, the highest $Tc^{onset}$ is about 5 K, while Tc descends slightly above 20 GPa.

To assure the drop observed in Figure 1(b) is indeed a superconducting transition, we further measured the resistance versus temperature at variant applied magnetic field(H). The evolutions of Tc at 18 GPa as a function of magnetic field are performed, as shown in Figure 2, with insets showing the change of Tc with H. It is obvious that Tc shifts toward lower temperature with the magnetic field increasing, indicating the transition is superconductivity in nature. According to the Werthamer-Helfand-Hohenberg (WHH) formula[31], $H_{C2}(0) = -0.691[dH_{C2}(T)/dT]_{T=Tc} * T_C$, the upper critical field $H_{C2}(0)$ is extrapolated to be 4.1 T for the onset of Tc , 3.4 T for the midpoint of Tc and 2.8 T for the zero point of Tc at 18 GPa with magnetic field H paralleling to *b* axis of $HfTe_5$ single crystal.

To determine the carrier density we further conducted Hall Effect measurement with a magnetic field H perpendicular to *ac* plane of $HfTe_5$ single crystal using Van der Bauer method. Carrier density increases almost three orders of magnitude with the pressure up to 9.8 GPa, as shown in Figure 3. It is visual that carrier density increases much faster above 5 GPa than that at lower pressure, which is consistent with the pressure where superconductivity appeared. In other world, the variations of Tc with pressure are closely connected with the change of carrier density. Combining with the change of calculated mobility with pressure, we supposed that $HfTe_5$ might transfer from the low pressure phase to the high pressure phase at around 5 GPa. The type of carrier is n-type in the whole range of pressure.

Raman spectroscopy is sensitive to local bond vibrations and symmetric broken, therefore it can provide evidence for the phase evolution of $HfTe_5$ under high pressure. Raman spectra (at 300K) of $HfTe_5$ at pressure and the evolution of the Raman vibration modes frequency as functions of pressure are shown in Figure 4. At 0.4GPa, six Raman peaks are corresponding to two $B_{2g}$ crystal modes of 68cm$^{-1}$ and 86cm$^{-1}$, four Ag modes of 116cm$^{-1}$, 120 cm$^{-1}$, 142cm$^{-1}$, and 168cm$^{-1}$, respectively[32]. With increasing pressure, two clear new Raman peaks appear at the pressure 3.7 GPa, which indicates the occurrence of a phase transition. Almost all phonon modes show blue shift as pressure increases up to 9.9 GPa, while the $B_{2g}$(1) mode of 68cm$^{-1}$ blue shifts firstly, and then shift toward lower wave number above 3.7 GPa as pressure further increase with a maximum of 74cm$^{-1}$. In fact, materials that exhibit resistivity change or superconductivity are often associated with structure distortion characterized by phonon softening subsequently resulting in either a structural or electronic structural phase transition[33-36]. Since the $B_{2g}$(1) mode is corresponding to the crystal structure and correlated to the $A_g$ mode of the pair of Te(II) atoms, it is assumed that there would be an electronic or structure phase transition above 3.7 GPa, corresponding to the appearance of superconductivity. Above 10.7 GPa, all the Raman peaks vanish or too weak to detect caused by a structural transition.

In order to check whether the pressure induced superconducting phase is caused by crystal structure phase transition, we performed crystal structure studies based on first-principle calculations of $HfTe_5$ at pressure up to 40 GPa. The enthalpies of the newly predicted stable phases, calculated at the high level of accuracy, are plotted as a function of pressure as shown in Figure 5. The ambient pressure *Cmcm* structure is the most stable phase up to 5 GPa, followed by a phase transition to monoclinic *C2/m* structure, which corresponds to the appearance of the superconductivity at 5.5 GPa in the transport measurement. Beyond 12 GPa, triclinic *P-1* structure becomes the most stable phase until 40 GPa. The crystal structure of *C2/m* and *P-1* were shown in Figure 5(b) and(c), respectively. In considerations of the transport and Raman experiments, the occurrence of superconductivity is possibly related to the transition from *Cmcm* to monoclinic *C2/m.* Comparable to our results of crystal structure prediction on $HfTe_5$, $ZrTe_5$ presents two structural transitions occurred around 6.0 GPa to *C2/m* structure and above 30 GPa to triclinic *P-1* structure, and the two new structures are responsible for the two superconducting phases[30].

In order to investigate the electronic structure evolution of $HfTe_5$, we also study the bulk state via first-principle calculations by taking into account SOC. Figure S2 shows that $HfTe_5$ is a weak topological insulator at 0GPa. With the pressure up to 10GPa and 20GPa, it transforms to metal, with complicated Fermi surface that coincides well with experimental results.

Referring to the results of electrical transport and structure predicted under high pressure, the phase diagram of $HfTe_5$ under pressure is shown in Figure 6. It can be seen that $HfTe_5$ remains the electronic state with topological character similar to that at ambient pressure below 5.5GPa. At the same time, the crystal structure remains the ambient pressure phase with space group *Cmcm*. The abnormal peak temperature of resistance decreases down to 40K firstly, and then rises up with increasing pressure. It reaches the highest value ~136K at 5.5GPa, above which superconductivity occurs. Accompanying with the appearance of superconductivity, a new crystal structure with the space group *C2/m* appears. Further increasing pressure, the bulk superconducting phase is stable in the pressure range up to 35GPa, with the highest $Tc^{onset}$~ 5K at 20GPa. But the transition to the second high pressure phase with the space group of *P-1* seems to have barely effect on evolution of Tc with pressure.

## Summary

In conclusion, combining experimental and theoretical investigations, we report the discovery of pressure induced superconductivity and related properties evolution in $HfTe_5$. The comprehensive studies on electronic, magnetic, vibration & structural properties of $HfTe_5$ single crystals under high pressure indicate that pressure induced transition from a TI to metal and the superconducting transition at 5.5GPa is correlated to structural transition from *Cmcm* to *C2/m*. Furthermore, a systematic phase diagram on crystal and electronic properties as a function of pressure for $HfTe_5$ is presented.

**Acknowledgements**

This research was supported by NSF&MOST of China through research projects.

## Figure Captions:

Figure 1 Electrical transport properties of $HfTe_5$ single crystal. (a) Temperature dependence of *ac* plane resistance at low pressure. (b) The *ac* plane resistance as a function of temperature at various pressures showing a superconducting transition at high pressure.

Figure 2 The superconducting transition of $HfTe_5$ with applied magnetic field H perpendicular to the *ac* plane of the $HfTe_5$ single crystal at 18GPa. The inset shows Tc evolution as function of magnetic field H.

Figure 3 Pressure tuned changes on Tc, carrier density and mobility in $HfTe_5$ at various temperatures(LPP & HPP indicate low pressure phase & high pressure phase, respectively).

Figure 4 (a) Raman spectra of $HfTe_5$ at pressure; (b) Raman shift evolution with pressure.

Figure 5 (a)Calculated enthalpies per atom as functions of pressure up to 40 GPa. Crystal structure of *C2/m*(b) and *P-1* phase(c).

Figure 6. The phase diagram of $HfTe_5$ single crystal as function of pressure. Tp denotes the peak temperature of resistance anomaly. The red circle represents Tp/10. The blue square stands for the onset temperature of resistance drop. The yellow region corresponds to semimetal phase till 5GPa. Above 5 GPa, the blue area indicates superconducting phase.

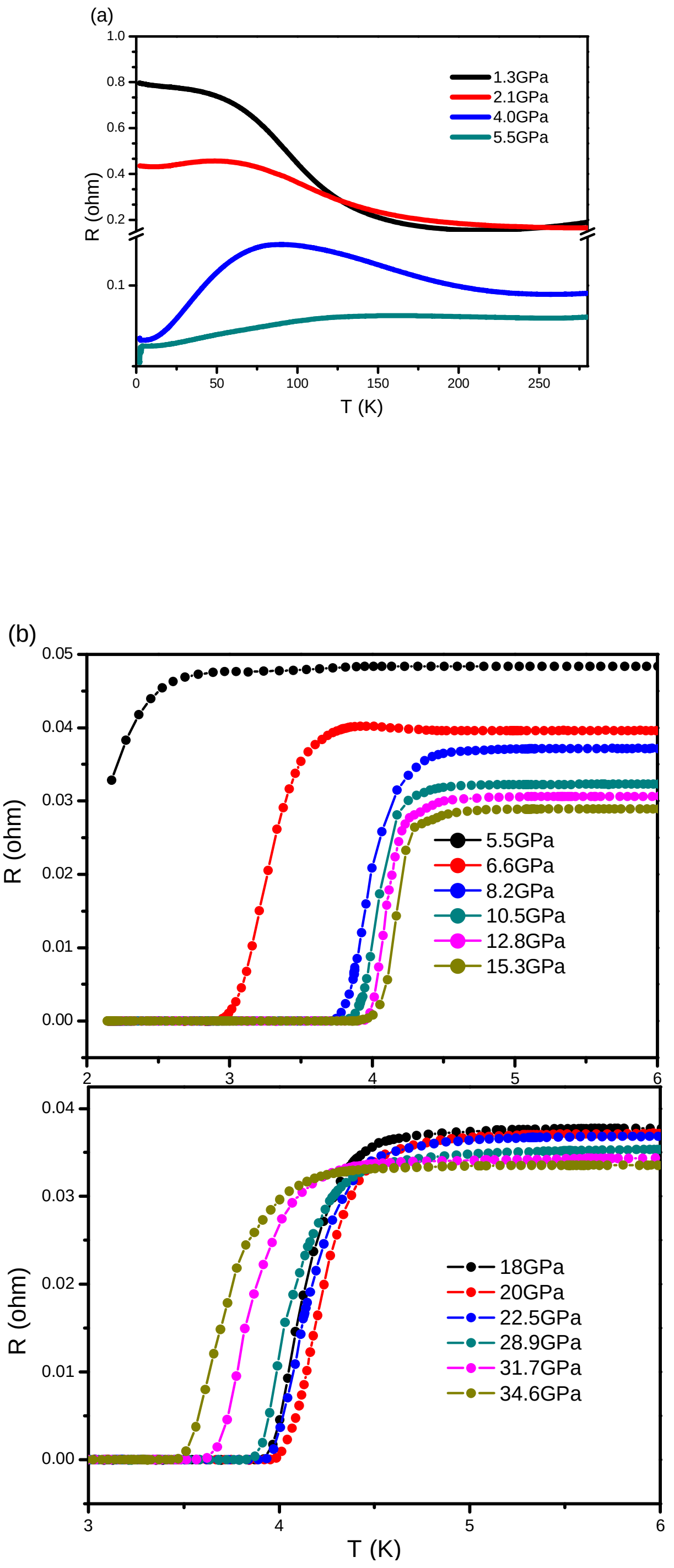

# Figure 1

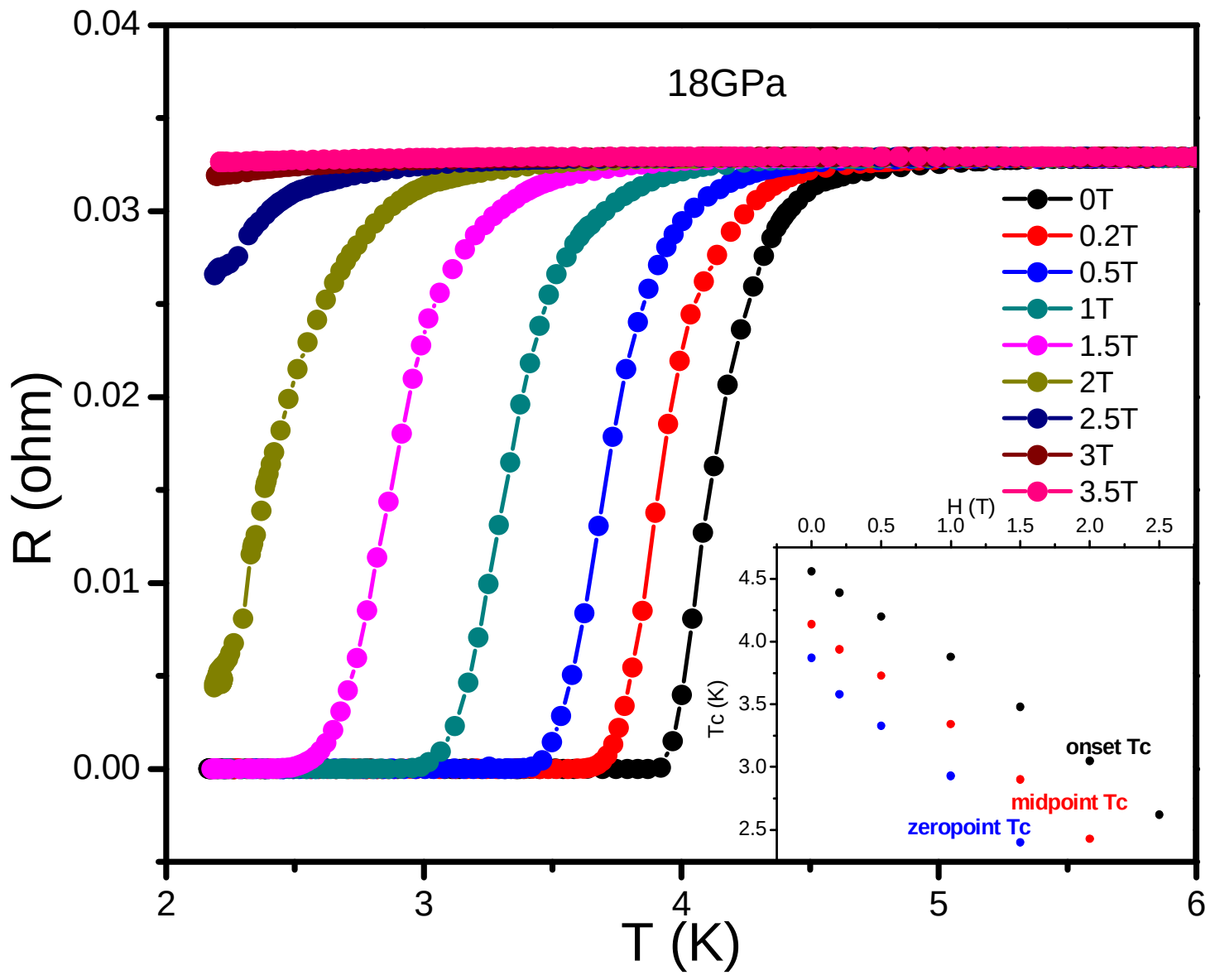

**Figure 2**

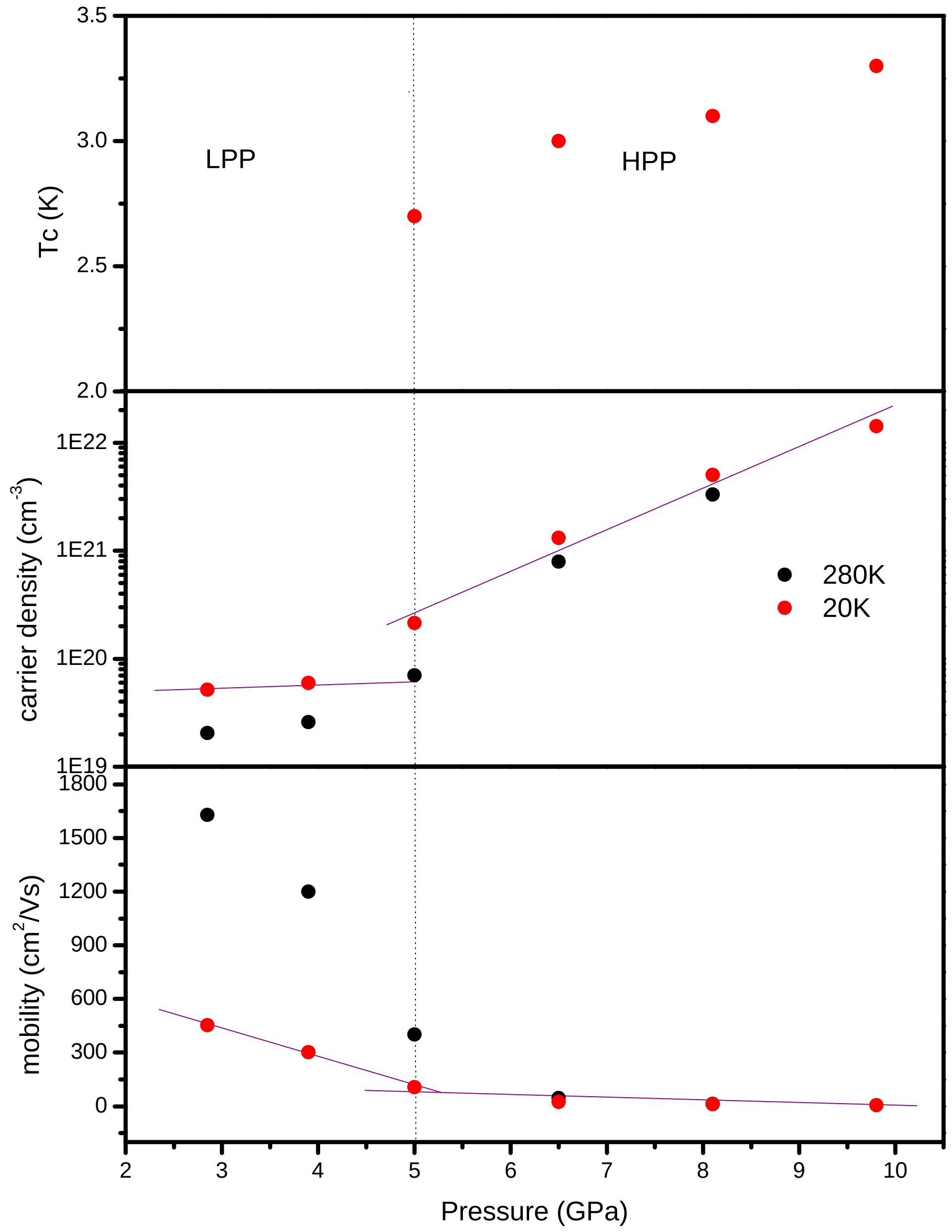

**Figure 3**

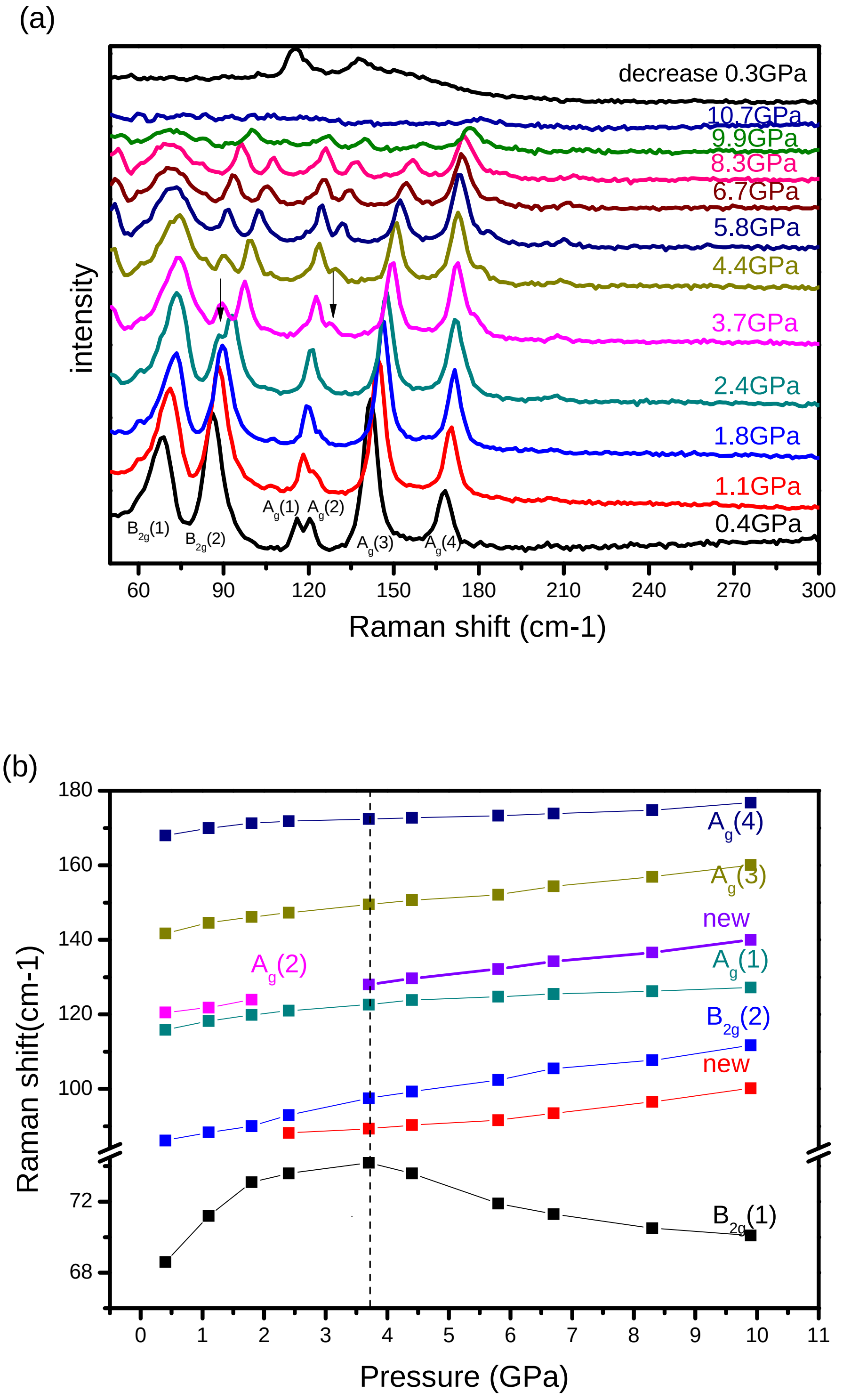

**Figure 4**

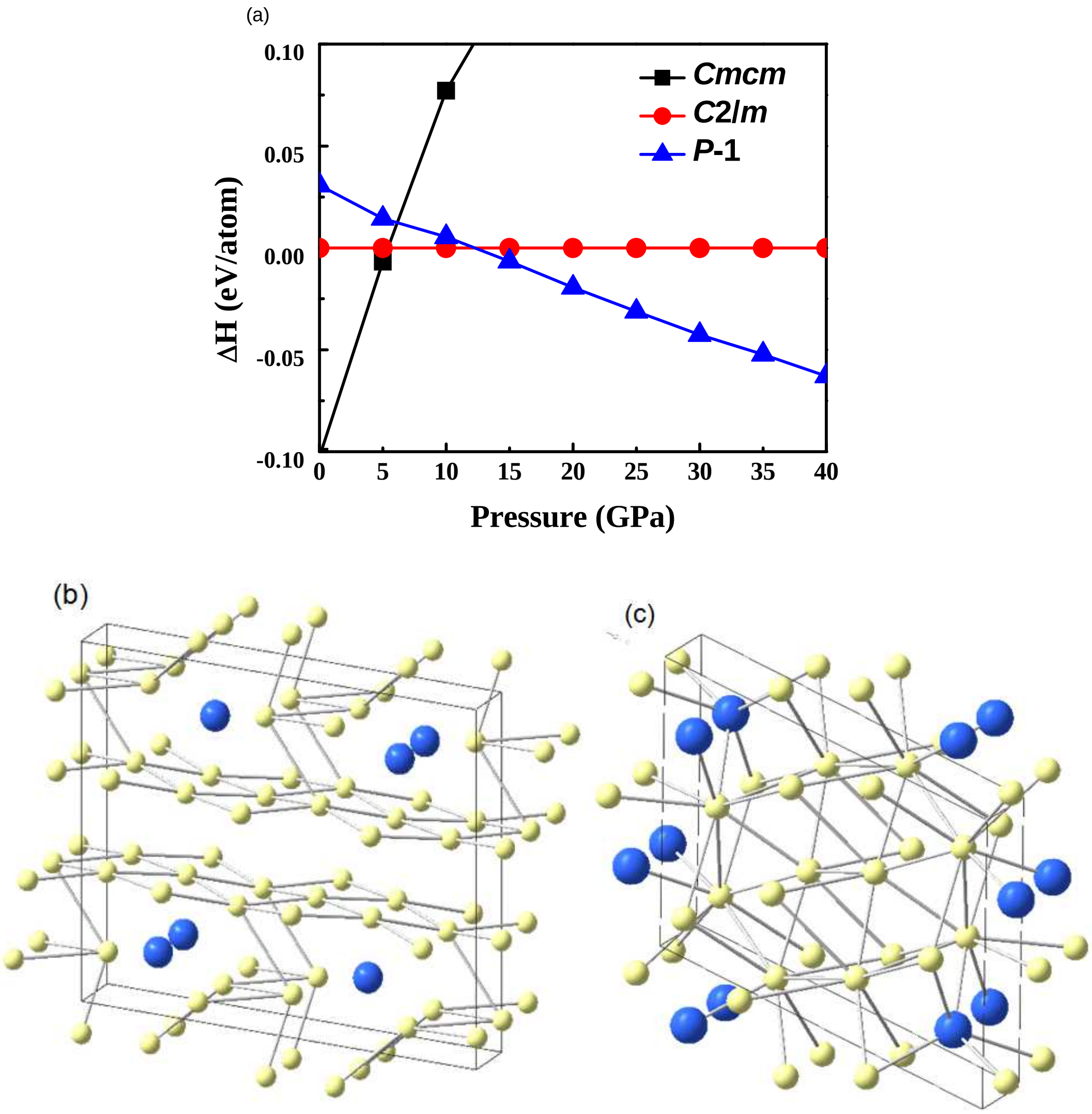

# Figure 5

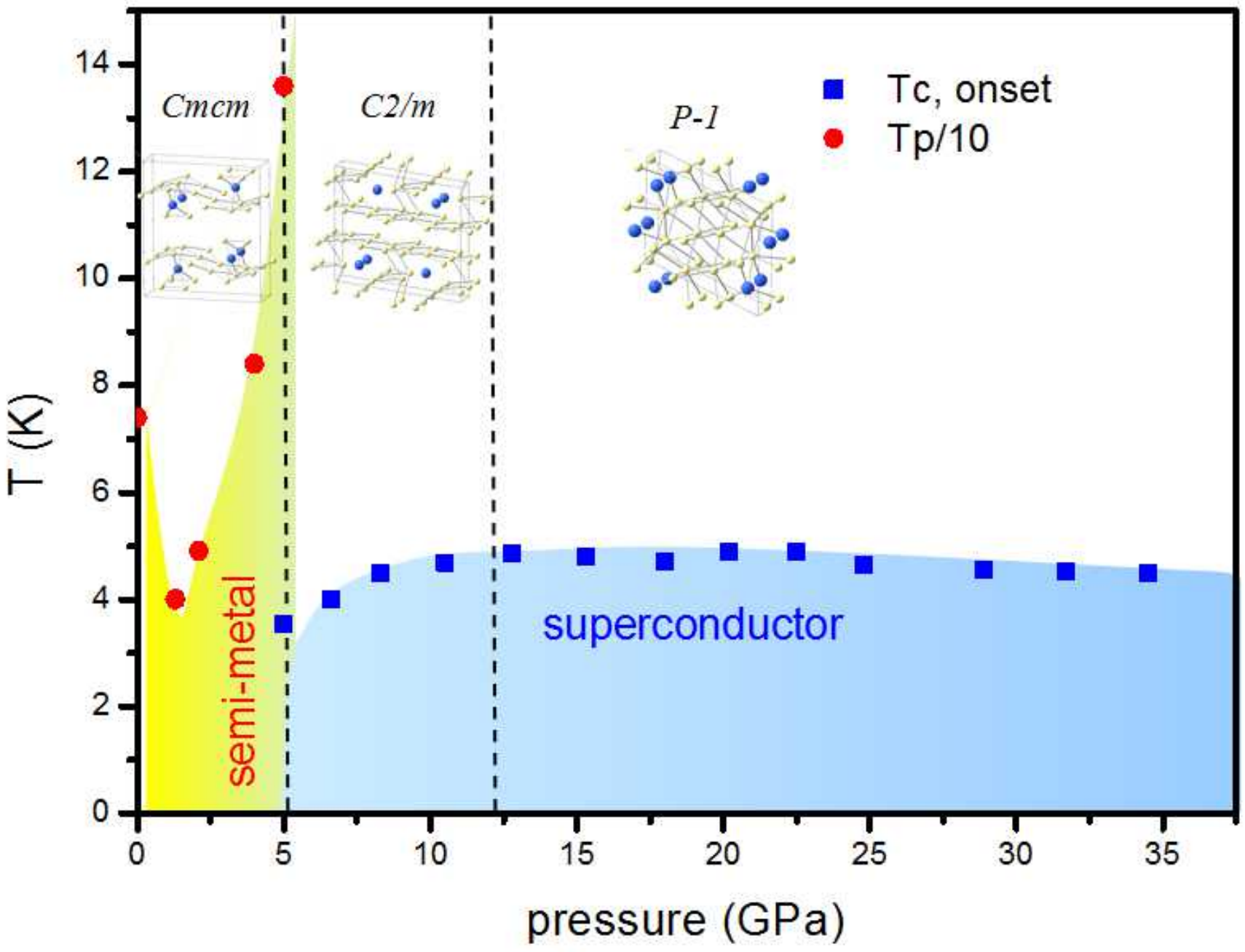

**Figure 6**

# Supplemental Materials

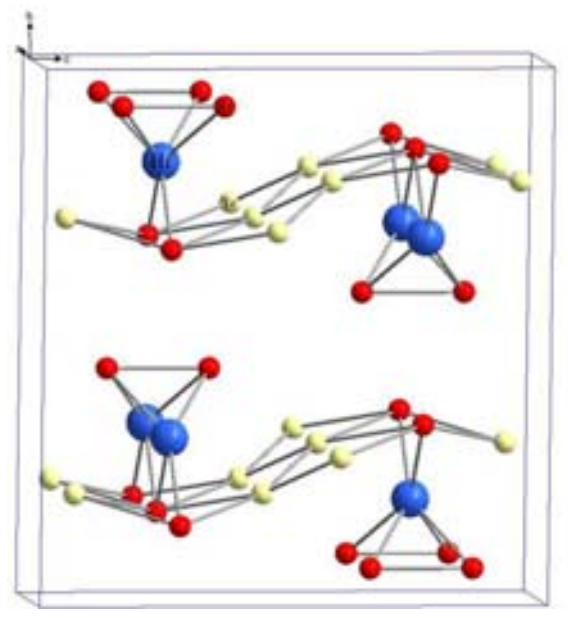

**Figure S1**: Schematic drawing of the crystallography of $HfTe_5$, where blue balls stand for Hf, red and yellow balls for different type of Te.

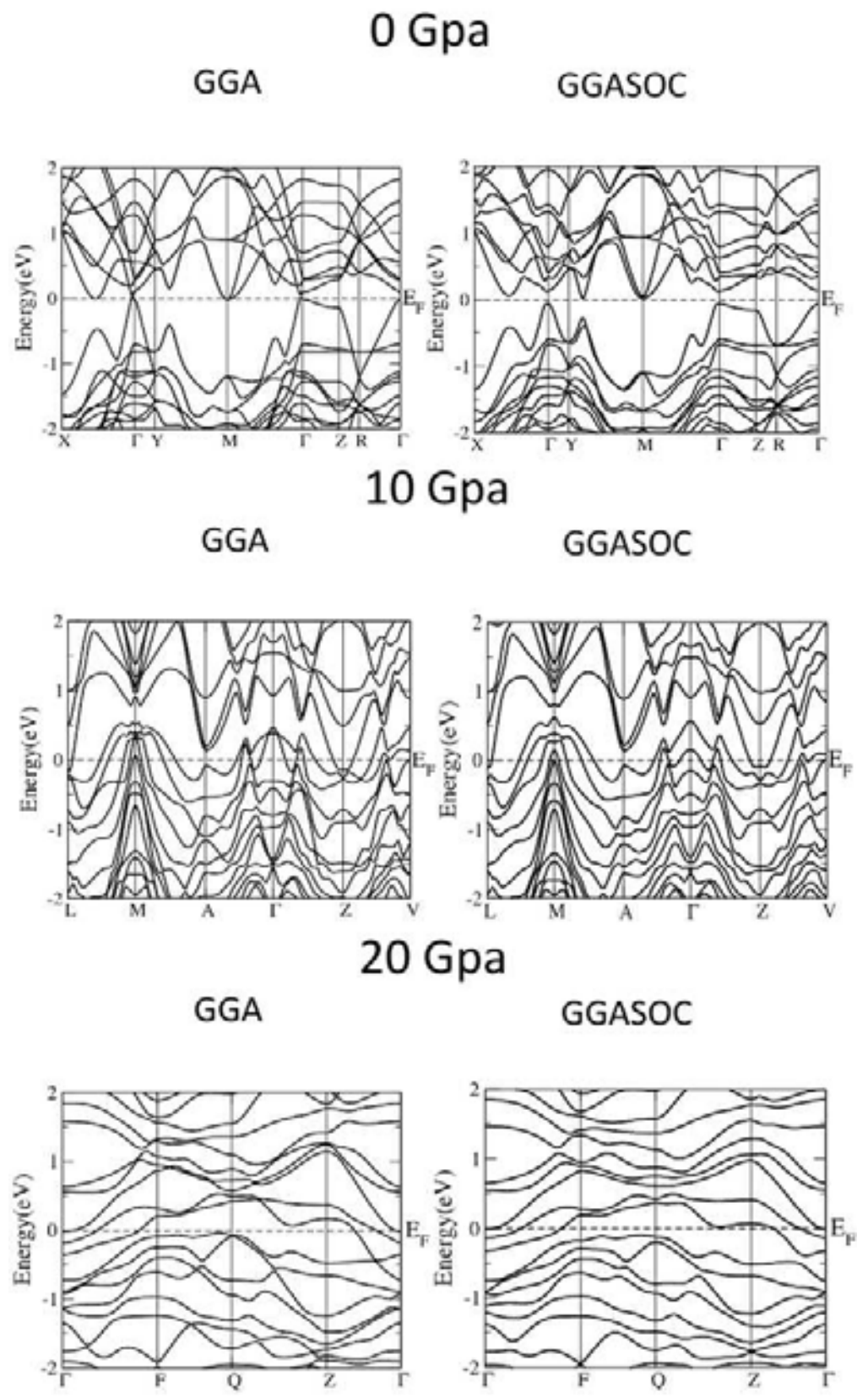

**Figure S2**: The calculated bond structure of $HfTe_5$ without and with SOC under 0GPa, 10GPa and 20GPa.